# Novel Cascaded Ultra Bright Pulsed Source of Polarization Entangled Photons

G. Bitton, W.P. Grice, J. Moreau[1], and L. Zhang[2]
Center for Engineering Science Advanced Research,
Computer Science and Mathematics Division, Oak Ridge National Laboratory,
Oak Ridge, TN

## Abstract

A new ultra bright pulsed source of polarization entangled photons has been realized using type-II phase matching in spontaneous parametric down conversion process in two cascaded crystals. The optical axes of the crystals are aligned in such a way that the extraordinarily (ordinarily) polarized cone from one crystal overlaps with the ordinarily (extraordinarily) polarized cone from the second crystal. This spatial overlapping removes the association between the polarization and the output angle of the photons that exist in a single type-II down conversion process. Hence, entanglement of photons originating from any point on the output cones is possible if a suitable optical delay line is used. This delay line is particularly simple and easy to implement.

[1] Ecole supérieure de Physique et de Chimie Industrielle de la Ville de Paris, 75005 Paris, France.

[2] Agiltron Corporation, 20 Arbor Lane Winchester, MA 01890.

## 1. Introduction

Entangled photon states generated by spontaneous parametric down conversion (SPDC) in non-linear crystals have been extensively used to test basic concepts in quantum mechanics [1]. In addition, entangled photon sources are used in an increasing number of applications, such as quantum teleportation [2], quantum communication [3], quantum cryptography [4], etc. Substantial effort is currently invested in improving the quality of those sources, both in terms of the total number of entangled states produced by the source and the degree of the entanglement of those states.

SPDC can be used to generate entangled photon states using type-I or type-II phase matching. In type-I phase matching, the two paired down converted photons have parallel polarizations, and they are emitted along concentric cones around the direction of the pump beam. In type-II phase matching, the two paired down converted photons are orthogonally polarized in the ordinary and extraordinary directions of the non-linear crystal. Those differently polarized photons are emitted along two different cones with cone axes at opposite sides of the pump beam and where the separation between the cones is determined by the cut-angle of the crystal. One of the first realizations of a bright, entangled photon sources made use of a type-II SPDC process in a degenerate non-collinear configuration [5] using a $\beta$-$BaB_2O_4$ (BBO) crystal. The cut-angle of the crystal was chosen to allow some overlapping of the extraordinarily polarized cone with the ordinarily polarized cone. Photon pairs emerging along the overlapping areas can be used to form polarization entangled states because their polarizations are not known *a-priori*. Those states can be simultaneously entangled in polarization, momentum direction, and energy. High pair production rates along with polarization interference visibilities exceeding 95% were reported. However, the brightness of type-II entangled photon sources is, in general, not optimized. Only a small fraction of the produced photon pairs can be used to form entangled states.

A significant increase in the entangled photon flux was achieved using a different setup, in which two type-I crystals are arranged in cascade with their optical axes perpendicular to each other [6]. By pumping with two orthogonally polarized pump beams (or one beam at $45^0$ with respect to the crystals optical axes) one can generate, for example, horizontally polarized photons from the first crystal and vertically polarized photons from the second. These two down conversion processes are mutually coherent (the pump beam is continuous with long coherence length), and bi-photon amplitudes from the two crystals can be used to form entangled states. The advantage of this source is that entangled states can be found anywhere on the output cones.

The entangled photon sources described so far are pumped by continuous laser beams with relatively long coherence times. However, considerable interest exists in entangled photon sources pumped by pulsed lasers for applications where timing of events is critical [7]. In a recent report, a cascaded type-I source operated in pulsed mode was described [8,9]. This source has a similar configuration to the cascade type-I source described above. Since it is operated in pulsed mode, a delay line must be used to allow overlapping in time of photon pairs created in the two different crystals. The pulse of the pump beam serves as a clock and allows us, *in principle*, to determine in which of the two crystals the detected pair was created. Such distinguishing information destroys the entanglement. Using a birefringent delay line, photons created in the first crystal can be delayed and overlapped in time with photons created in the second crystal. With 10 nm interference filters placed in front of the detectors, and a simple delay line consisting of a series of quartz plates, space-time interference visibility of more than 75% was observed [10] in a co-linear configuration where degenerate down converted photons propagate along the pump beam direction. For the more general (non-collinear) case the delay time between photons created in the first crystal and photons created in the second crystal will be angle dependent and an effective delay line will be more difficult to realize.

A special interest exists in type-II entangled photon sources. The two photons created in type-II SPDC are orthogonally polarized and hence it is easier to produce all four polarization-entangled Bell states. However, the visibility of type-II entangled photons sources operated in pulsed mode is limited. The visibility depends strongly on the thickness of the nonlinear crystal and the temporal and spectral widths of the pump beam [11,12,13,14,15]. Due to the birefringent nature of the crystal, the down converted photons travel inside the crystal at different speeds. Their relative arrival times with respect to each other and with respect to the pump pulse contain information about the polarizations of the individual photons and the location in the crystal where the pair was generated. This distinguishing information results in a reduced visibility, an effect which increases with crystal thickness. A partial solution to this problem is to use a birefringent delay line to introduce relative delays between orthogonally polarized photons. This removes some, but not all, of the distinguishing information. Further improvements may be realized through the use of a thin non-linear crystal [15,16] or narrow-band filters [14] which increase the coherence time. However, both methods reduce the available entangled photon flux reaching the detectors.

In this paper, we report on a new ultra-bright source of entangled photons operated in a pulsed mode using type-II SPDC in two crystals arranged in cascade. The optical axes of the crystals are at angles ψ and -ψ (see Fig. 1) with respect to the pump beam direction. In such an architecture the extraordinarily (ordinarily) polarized cone of the first crystal and the ordinarily (extraordinarily) polarized cone of the second crystal overlap. This overlapping removes the association between the photons' polarizations and their spatial location over the entire area of the output cones. Hence, one can use any part of the output cones as a source for entangled states, providing that a suitable delay line is installed.

In order to understand the temporal relationships between the photons produced in the cascaded type-II source, it is instructive to review the single-crystal source. Neglecting, for the time being, the complications due to pulsed pumping, the two-photon state for the single-crystal configuration may be written as

$$|y\rangle = \frac{1}{\sqrt{2}}\left[\left|H_{A,\frac{1}{2}(t_p+t_o)}, V_{B,\frac{1}{2}(t_p+t_e)},\right\rangle + \left|V_{A,\frac{1}{2}(t_p+t_e)}, H_{B,\frac{1}{2}(t_p+t_o)},\right\rangle\right]. \quad (1)$$

The form of this expression makes it apparent that a photon pair may be created with the ordinarily polarized photon (H) emitted into either beam A or beam B, and with the extraordinarily polarized photon (V) emitted into the conjugate beam. Also included in this expression are subscripts describing the average times of emission relative to the time when the center of the pump pulse enters the crystal. The delay $t_p$ is the time required for the pump to travel the full length of the crystal, while $t_o$ and $t_e$ are the propagation times for the o- and e-polarized photons, respectively. Since the average times of emission correspond to down-conversion events occurring at the crystal center, a horizontally polarized photon, for example, will have an average emission time equal to $(t_{p+} t_o)/2$. This is the time required for the pump to travel to the crystal center and for the o-polarized photon to subsequently propagate to the exit face.

Actually, for a CW pump the absolute times are less relevant than the relative times of emission. Paired photons created near the output surface of the crystal have identical emission times while paired photons created near the entrance face of the crystal have the largest difference in emission times. On average, the faster photon precedes the slower by $|t_o\text{-}t_e|/2$. Thus, information about the creation of a particular photon pair is carried in the relative emission times. If polarization entanglement is to be observed with this type of source, this timing information must be eliminated. One way that this may be accomplished is to allow the photons to pass through birefringent elements in order to delay the faster photon in each beam with respect to the slower by $|t_o\text{-}t_e|/2$. Including these delays, the source may then be described as emitting one photon into each beam, with an average arrival time in each beam that is independent of polarization.

The cascaded configuration is similar to the single-crystal source, in that there are two means by which photons of different polarizations may be emitted into beams A and B. This time, however, the two processes correspond to down-conversion events in different crystals. Because the two events occur in different regions of the source, a particular photon’s average emission time will depend not only on its polarization, but also on whether it originated in crystal 1 or crystal 2. Once again neglecting the complications associated with pulsed pumping, the two-photon state in this case may be written as

$$|y\rangle = \frac{1}{\sqrt{2}}\left[\left|H_{A,\frac{1}{2}t_p+\frac{3}{2}t_o}, V_{B,\frac{1}{2}(t_p+t_e)+t_{e'}}\right\rangle + \left|V_{A,\frac{3}{2}t_p+\frac{1}{2}t_e}, H_{B,\frac{3}{2}t_p+\frac{1}{2}t_o}\right\rangle\right], \quad (2)$$

where the delays are referenced to the time when the center of the pump pulse enters the first crystal. Upon comparison to Eq. (1), it is seen that additional delays are included: since the pump pulse must pass through the first crystal before reaching the second, all down-conversion events in the second crystal are delayed by $t_p$; and o- and e- polarized photons produced in the first crystal are delayed by $t_o$ and $t_{e'}$, respectively, as they propagate through the second crystal. Here, $t_o$, $t_p$, $t_e$, *and* $t_{e'}$ are defined as the propagation times thorough one of the cascaded crystals. The prime is included because the e-polarized photon generated in the first crystal has a different group velocity in the second.

As with the single-crystal source, it is necessary to eliminate the correspondence between average emission time and polarization. For Beam A, this entails delaying the e-polarized photon with respect to the o-polarized photon by $t_A=(3t_o-\tau_e-2t_p)/2$. This delay is different than the corresponding delay for Beam B, which is given by $t_B=(t_o-\tau_e-2t_{e'}+2t_p)/2$. In general, the average times of emission are angle-dependent. This is due not only to the angle-dependent index of refraction for the e-polarized photons, but also to the additional crystal material encountered in the beams making larger angles with the pump beam. The optimal delay, therefore, also depends on angle, as is the case with the type-I cascaded configuration (in non-linear configuration) described above. For our type-II cascaded configuration, however, a simple constant delay is found to be satisfactory for all output angles. This can be seen in Fig. 2, in which the average arrival times are plotted for different parts of the output cones. The parameters used for the calculation are the same as in our experimental setup. The calculated arrival times include constant delays of 410 fs and 31 fs for the e1 and e2 photons, respectively. Although the average arrival times vary across the output cones, the arrival times are nearly identical for the two photons that may be found in a particular beam. This property does not depend on the thickness of the crystals.

## 2. Experiment

The experimental setup is shown in Fig. 3. An argon ion laser is used to pump a femtosecond mode-locked Ti:Sapphire laser, producing a 76 MHz train of optical pulses centered near 790 nm. A second harmonic generation process in a 7-mm thick $LiB_3O_5$ (LBO) crystal with 40% efficiency is used to convert the beam from 790 nm to 395 nm. The spectral bandwidth of the pulse in the UV is 1 nm and its average intensity was usually 200 mW. A pair of prisms is used for dispersion compensation and fundamental removal. The beam is then passed through two consecutive thin BBO crystals (1.07 mm each) in which the optical axes of the two crystals are at angles $\psi$ = 43.65 and - $\psi$ with respect to the direction of the pump beam, as described above.

The alignment of the source was verified via a scan of the output pattern from the crystal pair. The output pattern consisted of two intersecting circles, just as would be expected from a single type-II SPDC with a cut-angle of $43.65^o$ (see the insert in Fig. 2). This indicates that the output cones from the two crystals completely overlap. We have also scanned the polarized output pattern by inserting a polarizer in front of the scanning head. Regardless of the polarizer angle the shape of the output pattern was the same and the intensity was half the intensity recorded without the polarizer. As expected, both vertical and horizontal polarized photons reside on the upper and the lower cones.

In order to demonstrate the unique character of the cascaded source, we chose to study the space-time and polarization interference of photon pairs emerging from regions where the upper and lower cones do not overlap. Two variable circular irises located 20 cm after the crystals were used to select the two specific spatial directions marked A and B in the insert of Fig. 2. Quartz plates and wedges placed after the irises serve as continuously adjustable birefringent delay lines. Polarizing beam splitters preceded by half wave plates were used as polarization analyzers (PA) and broadband spectral filters (SF) with cut-off wavelength at 715 nm were used to reduce the background noise. The coincidence rate at the two single-photon detectors (EG&G SPCM-AQR-14) was measured for various settings of the birefringent delays and for various orientations of the half wave plates.

The normalized coincidence rate for this set-up is calculated to be

$$R_c(q_A, q_B, t_A, t_B) = \frac{1}{2}\Big\{\left[Cos(q_A)Sin(q_B)\right]^2 + \left[Cos(q_B)Sin(q_A)\right]^2 + \sqrt{8p}\,Cos(q_B) \times Sin(q_B)Cos(q_A)Sin(q_A)Cos\left[v\,(t_A - t_B) + f_0\right]\frac{V(t_A, t_B)\,Rect(t_A, t_B)}{s\,(2t_p - t_o - t_e)}\Big\} \quad , (3)$$

where $\theta_A$ and $\theta_B$ are the polarization analyzer settings at detectors A and B and where $\tau_A$ and $\tau_B$ are the birefringent delays in arms A and B. As can be seen from this equation, the coincidence rate depends both on the angles of the polarizers and on the delay times. Hence, both polarization interference and space-time interference will be observed. The space-time interference is composed of a fast oscillating component given by $Cos[\,\phi_0 + \varpi(\,\tau_A - \tau_B)]$ and a slowly varying envelope given by $V(\tau_A\,,\,\tau_B)$. For a pulsed pump with a spectrum proportional to $I(\omega)=exp[-2(\omega-\varpi)^2/\sigma^2]$, the function $V(\tau_A\,,\,\tau_B)$ is

$$V(t_1, t_2) = Erf\left\{\frac{s}{4\sqrt{2}}\left[t_A - t_B + 4t_p - 2t_o - t_e - t_{e'} - \frac{2t_p - t_o - t_e}{t_o - t_e}\left|2t_o - t_e - t_{e'} - t_A - t_B\right|\right]\right\} - Erf\left\{\frac{s}{4\sqrt{2}}\left[t_A - t_B + t_e - t_{e'} + \frac{2t_p - t_o - t_e}{t_o - t_e}\left|2t_o - t_e - t_{e'} - t_A - t_B\right|\right]\right\} \quad . \quad (4)$$

$Rect(\tau_A,\tau_B)$ equals one for $t_o-\tau_e < t_A + t_B < 3t_o-\tau_e-t_{e'}$ and zero otherwise. The above expressions include the effects of group velocity walk-off, but neglect higher order dispersion [11,13].

The amplitude of the space-time interference depends on the angles of the polarizers. The interference term is largest when $\theta_A$ and $\theta_B$ are $\pm\pi/4$ and is zero for $\theta_A$ or $\theta_B$ equal $\pm\pi/2$ or 0. For given settings of the polarization analyzers, the visibility of the space-time interference is largest when $V(\tau_A, \tau_B)$ is maximized. As expected, this occurs when $t_A=(3t_o-\tau_e-2t_p)/2$ and $t_B=(t_o-\tau_e-2t_{e'}+2t_p)/2$. The calculated coincidence rate is shown in Fig. 4 as a function of the delay time in path B. For this plot, $\theta_A=\theta_B=\pi/4$, the delay in path A is constant and equal to $t_A=(3t_o-\tau_e-2t_p)/2$ and $\sigma$ was chosen to match experimental conditions. The oscillatory behavior and slowly varying envelope are evident in this plot. The insert shows the interference pattern on a smaller scale for $\tau_B \approx (t_o-\tau_e-2t_{e'}+2t_p)/2$. The maximum visibility of the interference pattern in Fig. 4 is 86%. The reason that it does not reach 100% is related to the pulsed nature of the pump and can be attributed to residual timing information that can not be eliminated with delay lines. As can be seen in Eq. (4), the maximum visibility attains its highest value in the limit $\sigma \rightarrow 0$, i.e., in the limit of a monochromatic pump.

With $\theta_A=\theta_B=\pi/4$, coincidence counts were measured for various values of $\tau_A$ and $\tau_B$. The general procedure involved setting $\tau_A$ and $\tau_B$ and then measuring the coincidences as small changes were made to $\tau_B$. Enough data points were collected to map out several fringes. This measurement was repeated many times for different starting values of $\tau_A$ and $\tau_B$. Fig. 5a shows data for $\tau_A \approx 31$ fs (1 mm of Quartz) and $\tau_B \approx 440$ fs (14.2 mm of Quartz), which represent the delays yielding the highest space-time visibility, as determined experimentally. The data in Fig. 5a has a visibility of 74%, which is somewhat below the theoretically predicted maximum of 86%. Note that all the measurements were conducted with cut-off filters at 715 nm. Such broadband filters do not increase the visibility and serve only to reduce the background noise. The observed fringe spacing in all our measurements was found to be 2.7 fs, which is in good agreement with the expected spacing of 2.63 fs. The small deviation of 2.5% from the expected result can be attributed to small misalignments of the crystalline quartz wedges. The difference in the optimal delay time for path B between the calculated value (410 fs) and the experimentally observed value (440 fs) can be attributed to small differences in the actual cut angle and thickness of the crystals.

The space-time interference pattern depends also on the angles of the polarizers in front of the detectors. The coincidence counts were measured for different angles of the polarizers and the results are in accordance with Eq. (3). Consider Figure 5b where the coincidence rate as a function of the delay time in path B is shown. The angles of the polarizers are 0 and $\pi/2$. The delay lines in paths A and B were set near their optimal values, namely, 1 mm and 14 mm of quartz, respectively. As expected from Eq. (3) the observed visibility is nearly zero.

The visibility of the space-time interference was extracted from the coincidence data and is plotted in Fig. 6 as a function of $\tau_B$. For this plot $\theta_A$ and $\theta_B$ were once again set to $\pi/4$ and $\tau_A$ was set to 31 fs. The solid line is the expected result from Eq. (3) using the experimental parameters, except that the crystal thickness was set to 1.1 mm (instead of 1.07 mm) to better fit the data. As can be seen, the general shape of the two curves is similar. That the maximum observed visibility is lower than expected can be attributed to imperfections in the experimental system. We have found that assuming a small deviation in the polarizers induces a significant decrease in the expected visibility. Note, also, that the width of the experimental results curve is not broader than the theoretical curve. This indicates that indeed, the spectral filters used in the experiment can be considered as broadband filters since they do not increase the coherence time of the down converted photons.

In addition to space-time interference, the cascaded source also exhibits polarization interference, which is manifested as a variation in the coincidence rate as a function of the relative angles between the polarization analyzers. This is typically observed in the following way: one of the polarization analyzers is set to a particular angle and the coincidence rate is measured as the other analyzer is rotated. For a maximally entangled state, the visibility will be unity, independent of the setting of the first analyzer. This would be the case for the cascaded source when pumped by a monochromatic pump, as long as $\tau_A$ and $\tau_B$ were optimized. For a pulsed pump, however, the additional timing information reduces the visibility. [Of course, if one of the analyzers is set to 0 or $\pi/2$, the visibility will be 100%, regardless of the delays and of the properties of the pump. However this is not a property of the entanglement but of the experimental setup.] The visibility is smallest (representing the most exacting test of polarization entanglement) when the fixed analyzer is set to $\pi/4$. The visibility of the polarization interference in this case is predicted to be the same as the visibility of the space-time interference. This is evidenced in Fig. 7, which shows polarization interference patterns. The delay lines were kept constant around their optimal values, $\tau_A \approx 31$ fs, and $\tau_B \approx 440$ fs, the first polarization analyzer was then held fixed and coincidences were recorded as $\theta_B$ was varied. As expected, the observed polarization interference visibility is close to the space-time interference visibility observed in Figure 5a.

## 3. Discussion

Practical implementations of entangled-photon sources require robustness, ease of implementation, high visibility and brightness. The type-II cascade source is an advance in this direction, especially the increased brightness with respect to single type-II sources and the ease of implementing the required delay lines. It was shown that simple delay lines could be used to temporally overlap in time photons generated in the different crystals over the entire area of the output cones. This unique property of the cascaded type-II structure greatly simplifies an actual construction of ultra-bright entangled photon sources.

The cascaded source may also be useful in experiments involving two or more pairs of photons. Unlike single crystal configurations [17,18] multiple pairs of polarization-entangled photons may be generated in distinct paths with only a single pass of the pump beam.

The maximum expected theoretical visibility of the cascaded source for our experimental conditions is limited to 86%. This limitation is typical of SPDC pumped by a short pulse and is more pronounced in type-II SPDC than in type-I SPDC. To overcome this limitation one can choose crystals and wavelengths [11,13], which minimize this effect or use a specific experimental technique developed to overcome this limitation [19,20].

In conclusion, we have constructed a new type of pulsed source of polarization entangled photon states based on type-II SPDC in two cascaded crystals. Due to the special geometry of the two-crystal setup, entangled photon states are formed anywhere on the output cones, resulting in an ultra bright source of entangled photons. We have measured the polarization and time-space interferences as a function of the polarizer angles and the delay lines. The observed visibilities of the interference patterns were found to be as high as 74% when suitable delay lines were used. We have shown that the same delay lines could be used for any point on the output cones thus enabling a simple realization of an ultra-bright source based on the type-II SPDC cascade configuration.

## Acknowledgements

It is a pleasure to acknowledge the contributions of J. Barhen, V. Protopopescu, H.K. Liu, C. D'Helon and Y.H. Kim. Their advice and comments are greatly appreciated. This research was supported by the US Department of Energy, Office of Basic Energy Sciences under contract No. DE-AC05-00OR22725 with UT-Battelle, LLC.

---

# Figure captions

FIG. 1. Emission cones from a cascaded type-II SPDC for the degenerate case. The optical axes of the two crystals $OA_1$ and $OA_2$ are at angles $\psi$ and $-\psi$ where $\psi$ is the cut-angle of the crystals. The pump beam propagates along the Z direction.

FIG. 2. The calculated average emission times as function of the angle $\phi$, where $\phi$ is the angle between the x-axis and the projection of the photon k-vector on the x-y plane. The emission times of the photons depend on their polarizations and on the crystal in which they were down converted. The legends used in this figure are: 1e (1o) is an extraordinarily (ordinarily) polarized photon down converted in the first crystal; 2e (2o) is an e- (o-) polarized photon down converted in the second crystal. The photons 1e, and 2e are delayed by a fixed amount of time with respect to photons 1o and 2o (410 fs and 31 fs respectively). When this delay is used, the average arrival times of the pairs (1e,2o) and (1o,2e) are practically the same for all output angles. The insert in Figure 2a shows the values of the $\phi$ angles for different points on the output cones. The two shaded circles indicate the spatial directions used in this experiment.

FIG. 3. Experimental setup.

FIG. 4. The coincidence counts as function of the delay-time in path B as computed from Eq. (3). The delay-time in the other optical path is kept constant and equal to $\tau_a$. The angles of the polarizers in front of the detectors are $\pi/4$. The insert in the figure is a smaller scale of the interference pattern around $\tau_1=\tau_a$ $\tau_2=\tau_b$. As can be seen the fringe spacing is 2.63 fs.

FIG. 5. The coincidence counts as function of the delay-time in path B. The delay time in path A is 31 fs and the delay time in path B was changed around 440 fs. The angles of the polarization analyzers in (a) are $\pi/4$ and the observed visibility is 74%. The angles of the polarization analyzers in (b) are 0 and $\pi/2$ and the observed visibility is practically zero.

FIG. 6. The visibility of the space-time interference as function of the delay time in path B. The delay time in path A was kept constant and equal to 31 fs. The solid line is the visibility calculated from Eq. (3).

FIG. 7 The coincidence counts as function of the angles of one of the PA. The angle of the other PA was at $\pi/4$. The delay time in paths A was 31 fs. The delay time in path B was about 440 fs. The observed visibility is 73%.

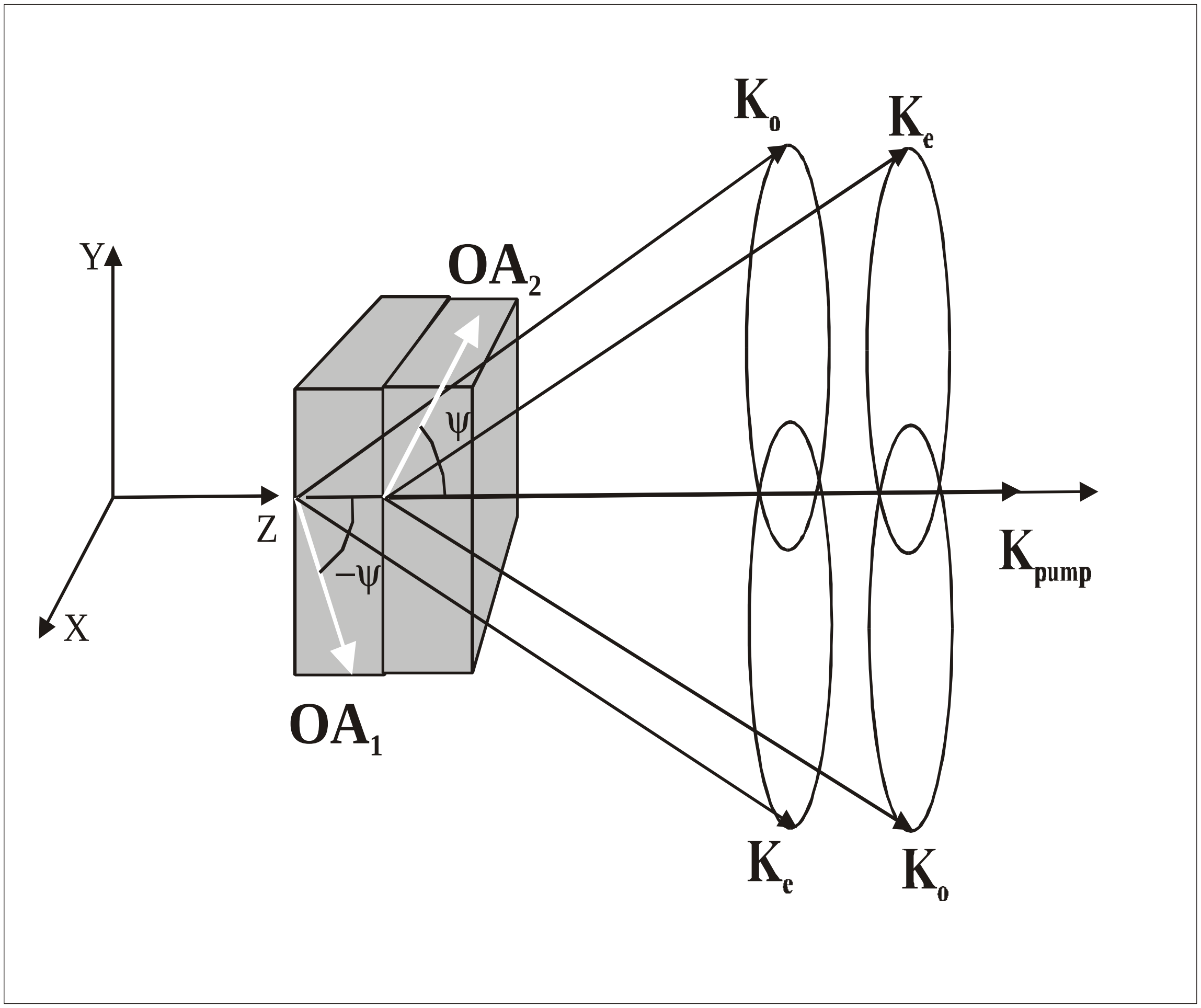

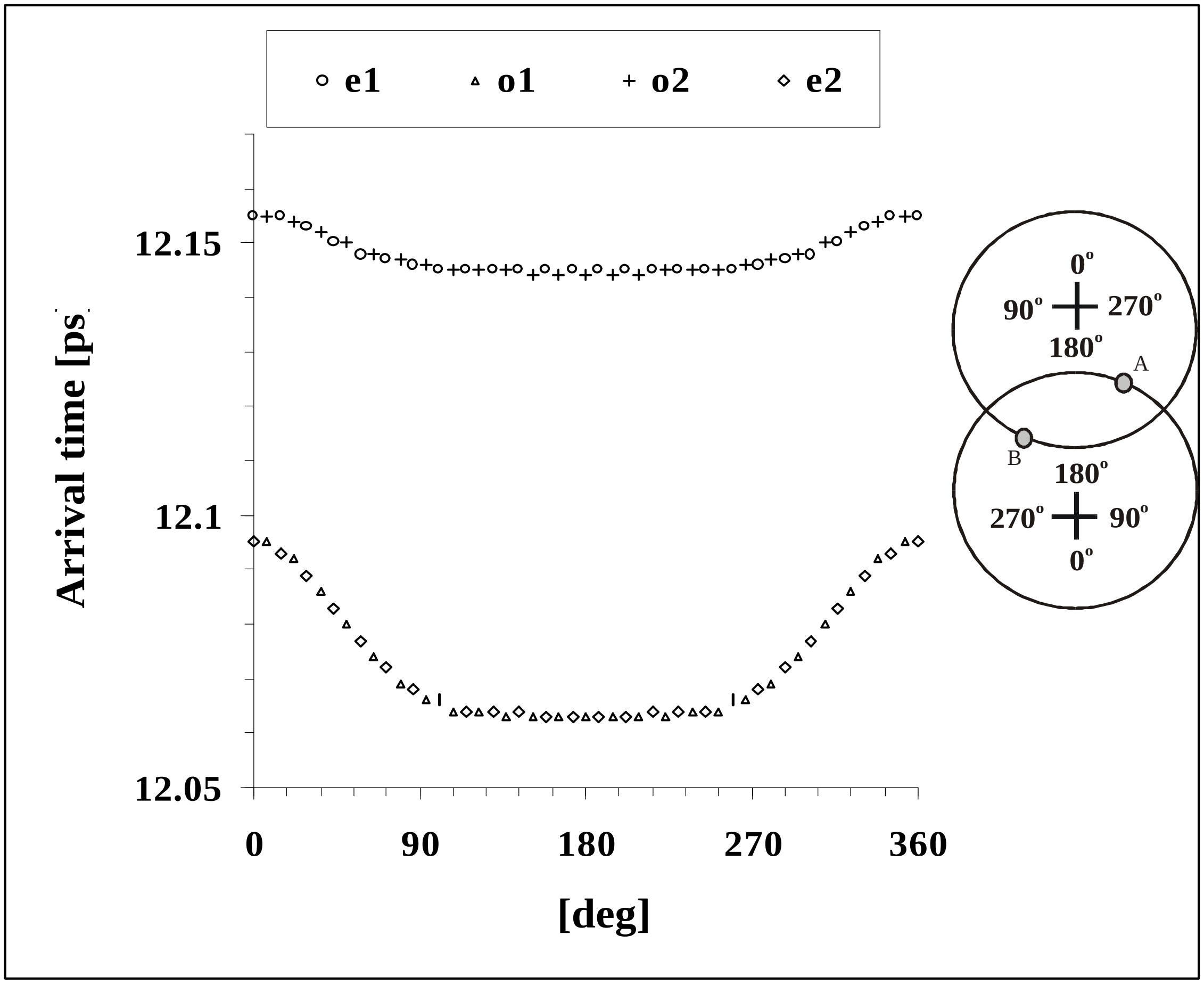

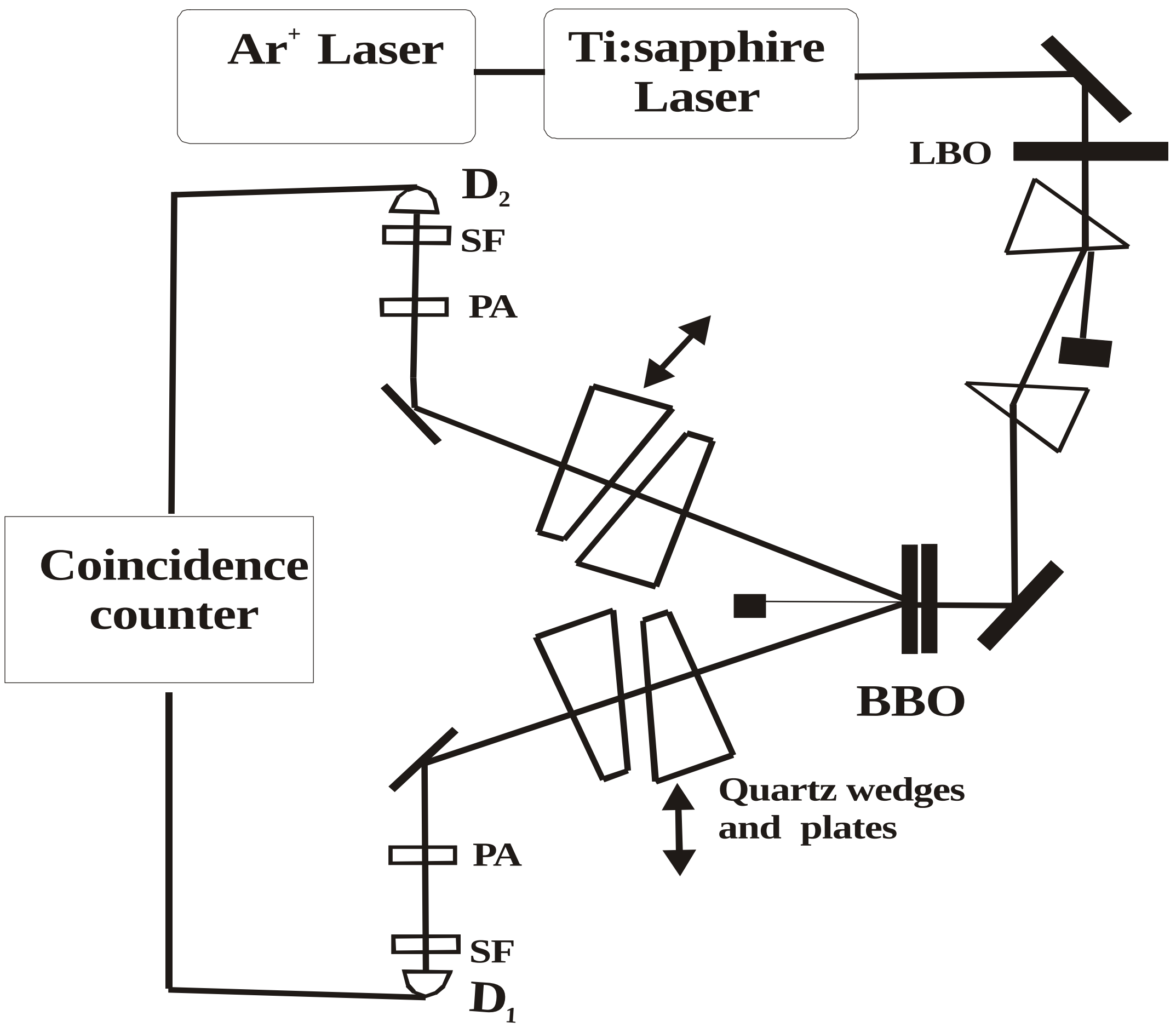

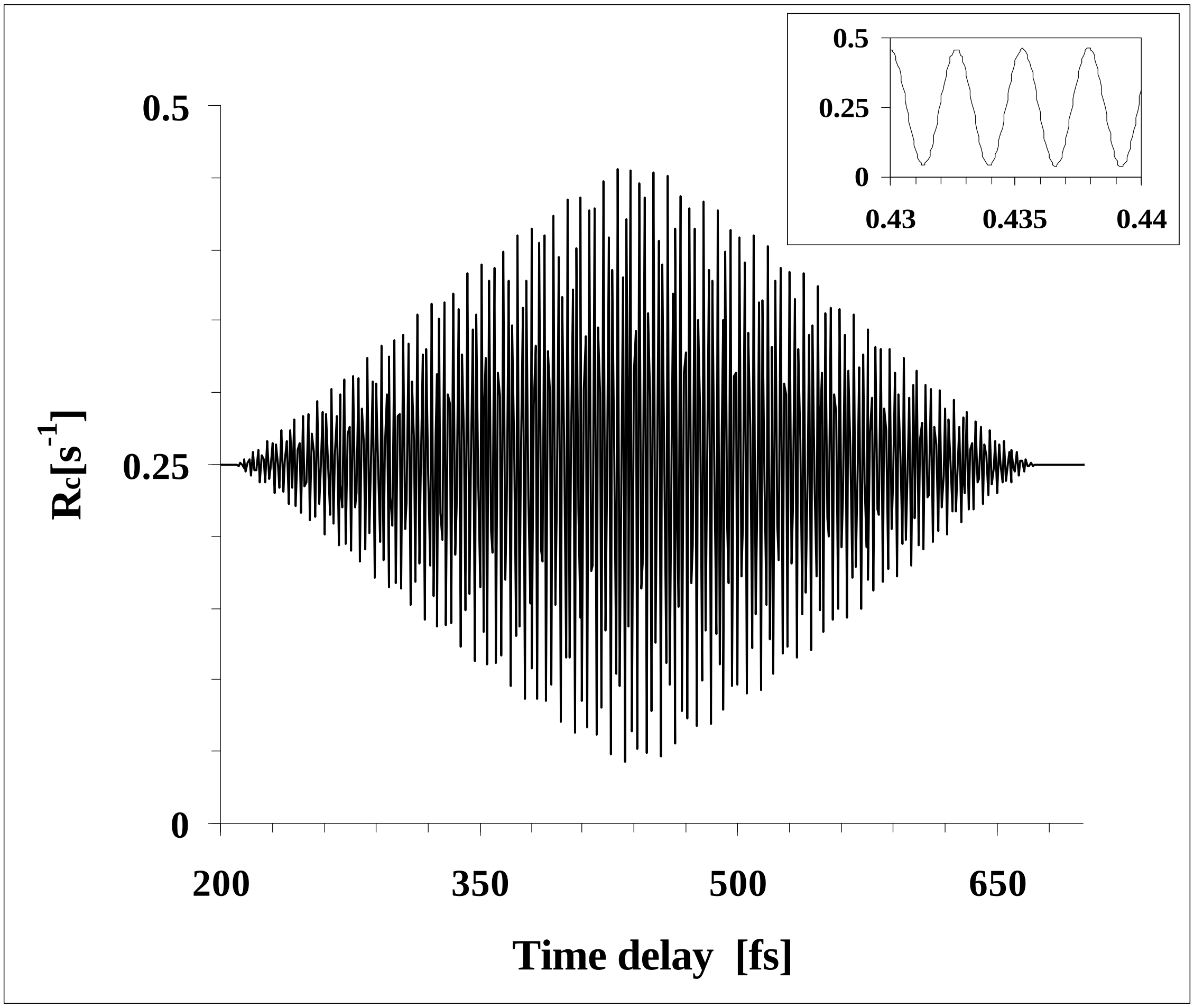

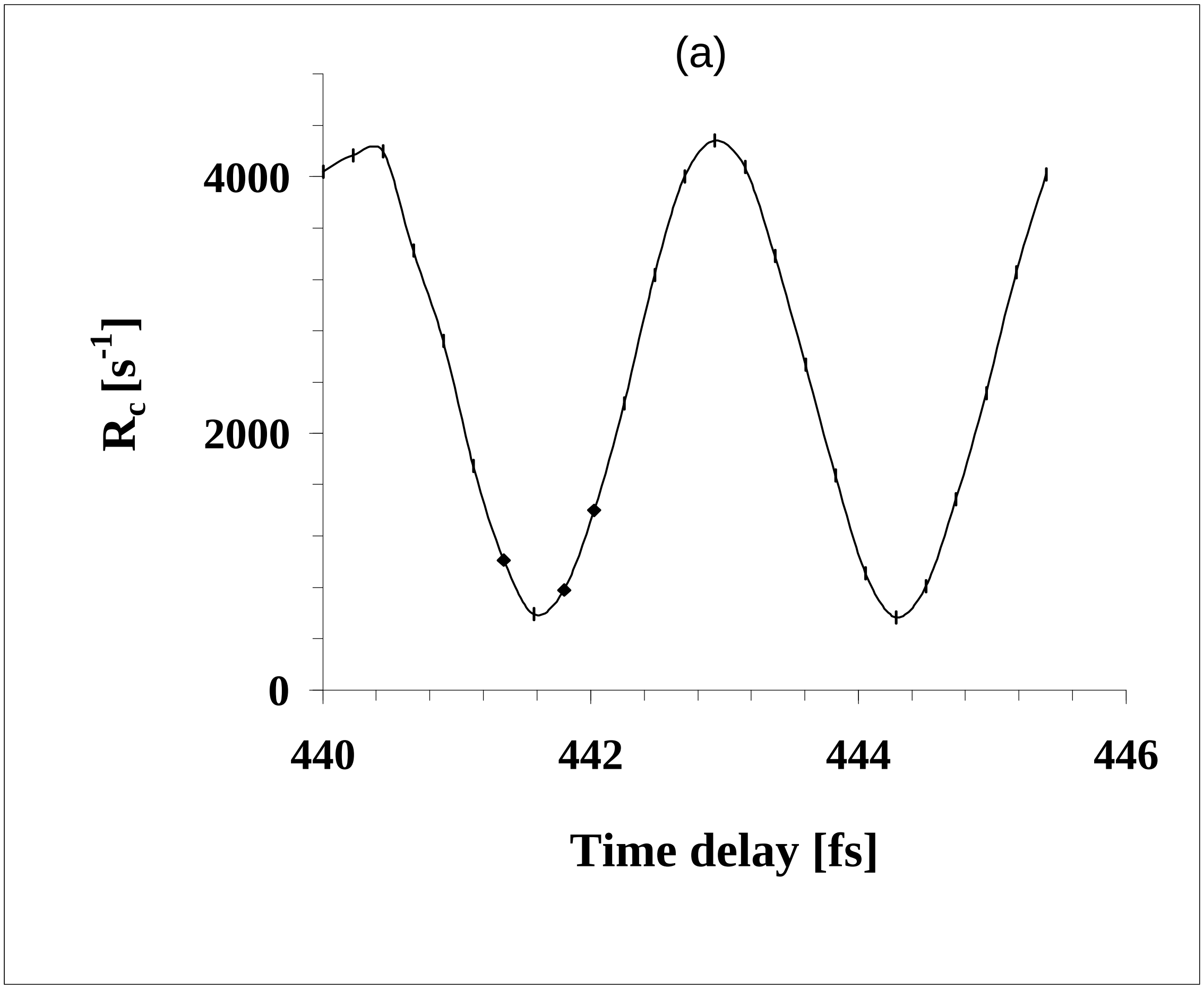

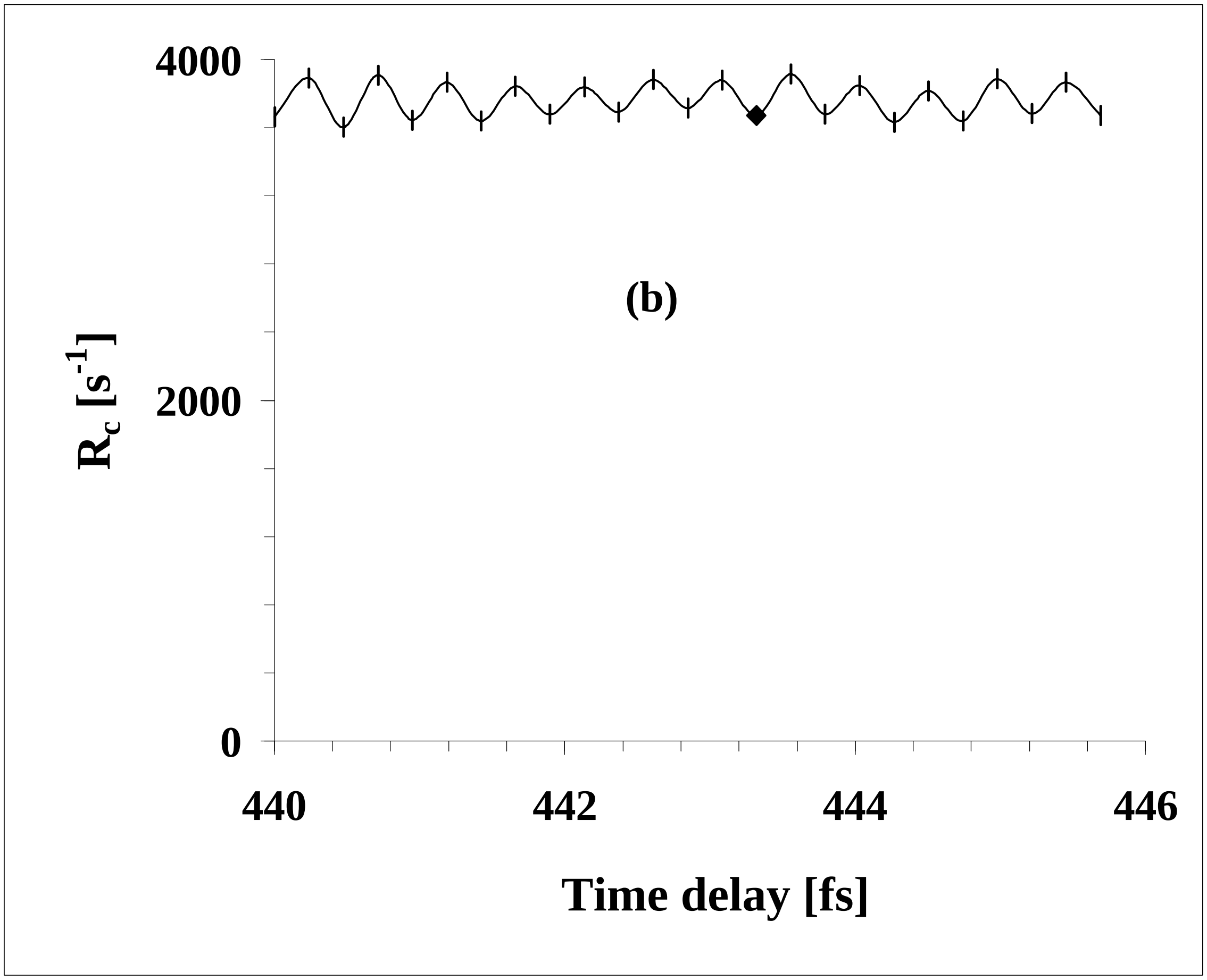

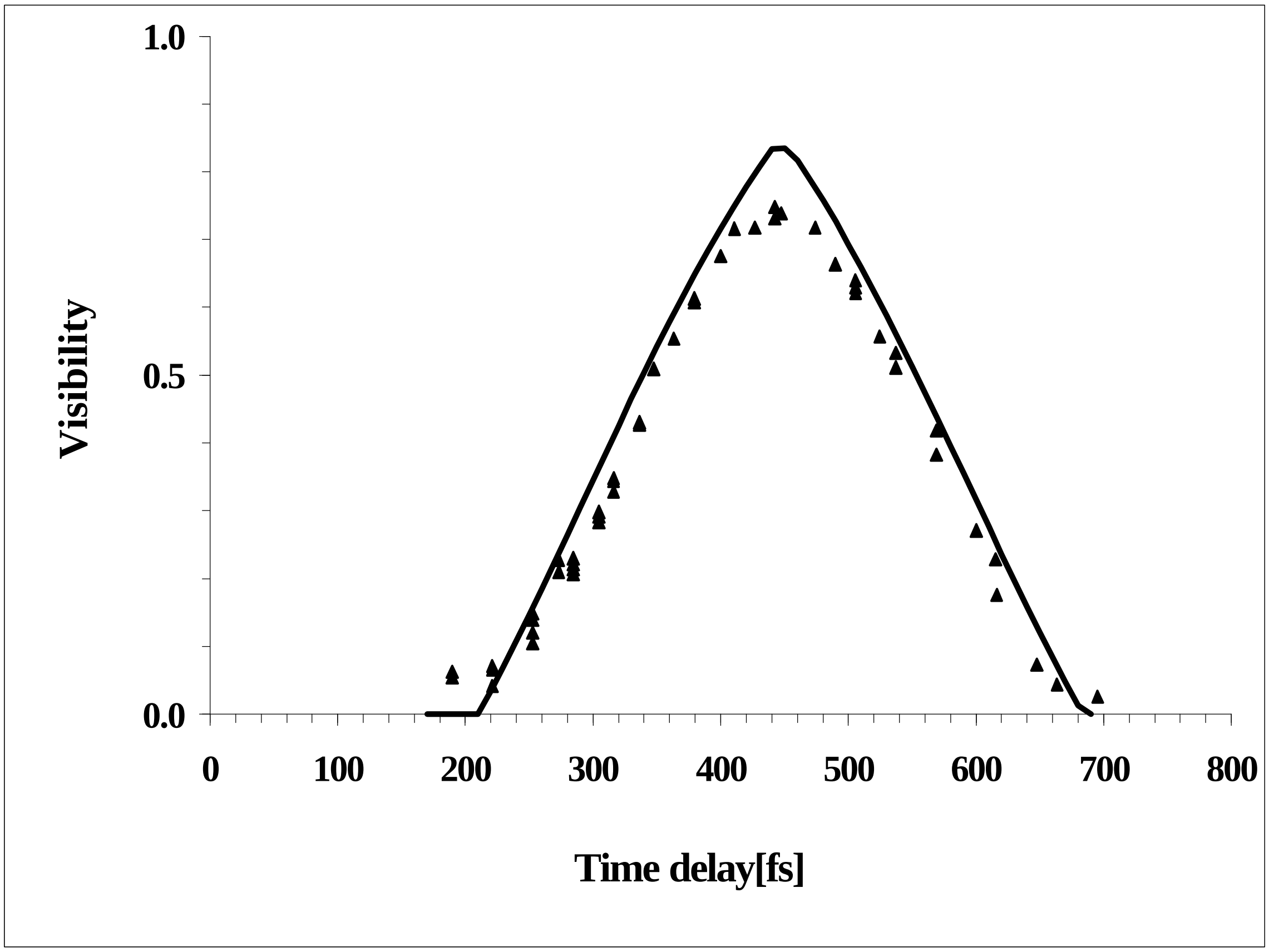

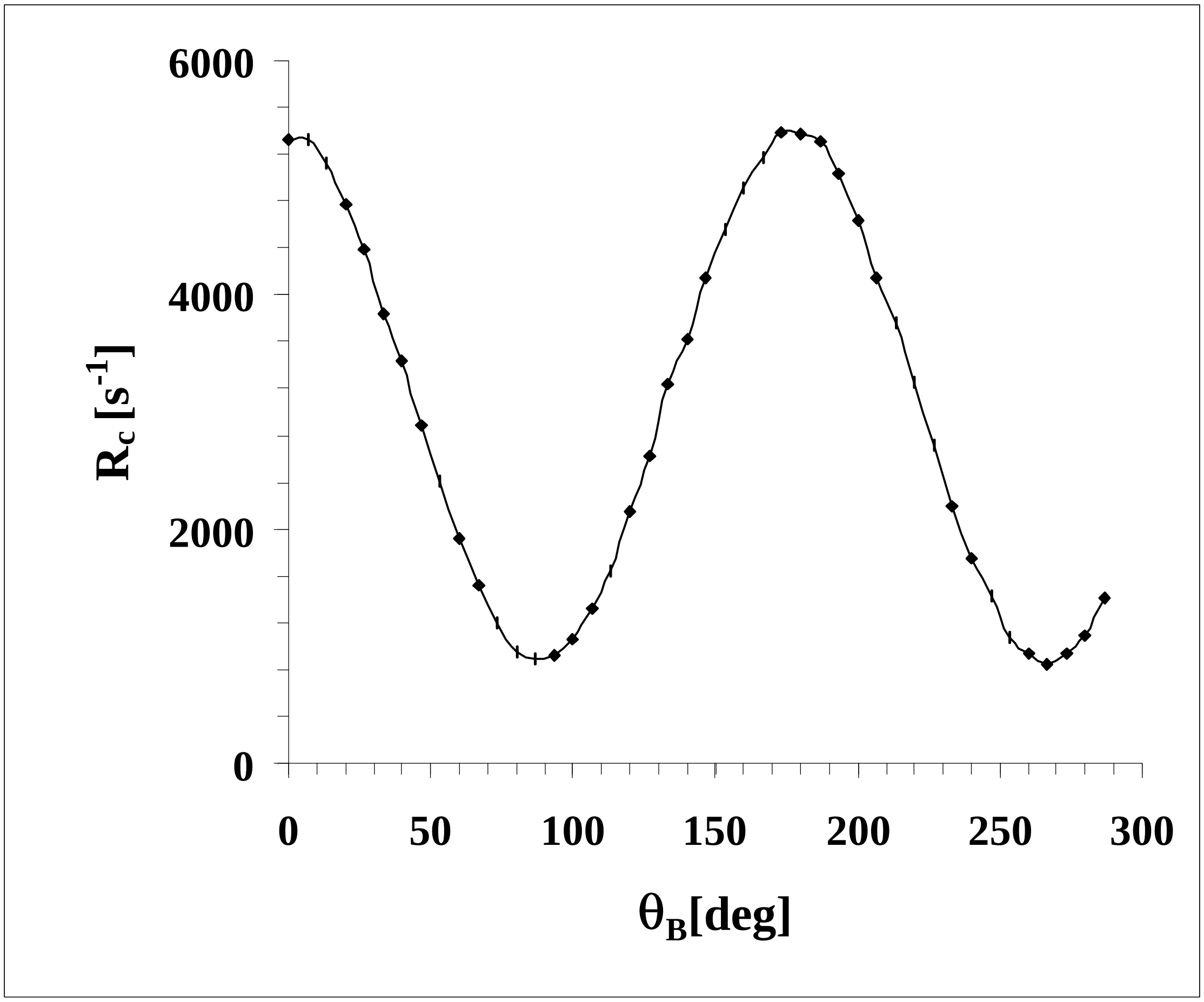